# Algorithmic metatheorems for decidable LTL model checking over infinite systems

Anthony Widjaja To and Leonid Libkin

LFCS, School of Informatics, University of Edinburgh
`anthony.w.to,libkin@ed.ac.uk`

**Abstract.** By algorithmic metatheorems for a model checking problem $P$ over infinite-state systems we mean generic results that can be used to infer decidability (possibly complexity) of $P$ not only over a specific class of infinite systems, but over a large family of classes of infinite systems. Such results normally start with a powerful formalism $F$ of infinite-state systems, over which $P$ is undecidable, and assert decidability when is restricted by means of an extra "semantic condition" $C$. We prove various algorithmic metatheorems for the problems of model checking LTL and its two common fragments $\text{LTL}(\mathbf{F}_s, \mathbf{G}_s)$ and $\text{LTL}_{\text{det}}$ over the expressive class of word/tree automatic transition systems, which are generated by synchronized finite-state transducers operating on finite words and trees. We present numerous applications, where we derive (in a unified manner) many known and previously unknown decidability and complexity results of model checking LTL and its fragments over specific classes of infinite-state systems including pushdown systems; prefix-recognizable systems; reversal-bounded counter systems with discrete clocks and a free counter; concurrent pushdown systems with a bounded number of context-switches; various subclasses of Petri nets; weakly extended PA-processes; and weakly extended ground-tree rewrite systems. In all cases, we are able to derive optimal (or near optimal) complexity. Finally, we pinpoint the exact locations in the arithmetic and analytic hierarchies of the problem of checking a relevant semantic condition and the LTL model checking problems over all word/tree automatic systems.

## 1 Introduction

The study of model checking over infinite-state systems is now an active research area. This can be justified by the plethora of real-world scenarios that can be more conveniently modeled using infinite-state systems rather than finite-state systems, e.g., those that typically arise as programs with unbounded data structures (including stacks, lists, and FIFO queues), numeric variables, and clocks. To make sense of the problem of verifying infinite-state systems, the systems under consideration need to have some finite representations, e.g., timed automata, pushdown automata, counter machines, Turing machines, and so forth. Unlike in the case of finite systems, model checking even the most basic properties, such as safety and liveness, is already undecidable over infinite-state systems in general. For this reason, one either adopts non-Turing-powerful formalisms which admit decidability or resorts to semi-algorithms for general formalisms. Examples of formalisms that admit certain decidable model checking problems include pushdown automata [9], higher-order pushdown automata [20], Petri nets [9], timed automata [3], lossy channel systems [1, 4], certain subclasses of counter machines [11, 21, 24], and classes of term/tree rewrite systems [9, 25, 31], to name a few. On the other hand, various notions of finite-state transducers on words/trees [2, 7, 12, 15, 29, 34] and certain extensions of counter machines [6, 11] have emerged as popular general (Turing-powerful) frameworks of infinite-state model checking, for some of which semi-algorithmic approaches to verifying basic model checking properties, including safety and liveness, have been proposed (e.g. see [2, 6, 8] and references therein).

The vast literature of infinite-state model checking in the past decade or so can be extremely daunting and easily obscure proof patterns that can be *reused with ease* to obtain model checking decidability results for seemingly unrelated formalisms of infinite-state systems. This issue motivates the study of *algorithmic metatheorems* for infinite-state model checking, which are generic results that can be used in a "plug-and-play" manner for inferring decidability of certain model checking tasks over *a large family* of formalisms of infinite-state systems, instead of doing so for *a single* formalism at a time. Of course the concept of algorithmic metatheorems is not new; the classical decidability of S2S can be viewed as such for MSO model-checking, due to its wide applicability via the method of interpretations. Other results of this nature include results on flat counter machines [11, 24], well-structured transition systems [4, 17], and the extension of the S2S result to Caucal hierarchy [33]. In the finite case, algorithmic metatheorems are used extensively to obtain good algorithmic bounds for evaluating logical formulae over finite structures [19].

In this paper we study generic algorithmic metatheorems for designing efficient algorithms for model checking LTL, together with two of its commonly considered fragments LTL($\mathbf{F}_s, \mathbf{G}_s$) [31, 32] and LTL$_{\text{det}}$ [27], over infinite-state systems. Our choice of logic is justified by the fact that LTL, LTL($\mathbf{F}_s, \mathbf{G}_s$), and LTL$_{\text{det}}$ can express frequently checked properties including safety and liveness, and that their model checking problem have been frequently studied in the setting of infinite-state systems (e.g. [8, 9, 22, 31]). We will use as our framework the expressive class of *word/tree automatic transition systems* [7, 15, 34], which are generated by *synchronized rational transducers* [12, 15] over finite words and finite ranked trees. Such systems subsume many important decidable formalisms, including many which we previously mentioned and others, and still possess desirable closure and decidability properties (e.g. see [7, 12]), many of which are not satisfied by the general class of rational transducers on words [29]. Since verifying safety and liveness are in general undecidable over automatic transition systems [7], we will study various "semantic restrictions" for ensuring decidability of LTL, LTL($\mathbf{F}_s, \mathbf{G}_s$), or LTL$_{\text{det}}$, without unnecessarily sacrificing *applicability* and *algorithmic efficiency*.

*Contributions*    We identify semantic conditions on word/tree automatic transition systems that let us conclude decidability (and complexity) of model-checking. We start with a condition **(C1)** stating that the reachability relation is effectively computable and given by a synchronized rational word/tree transducer. There are many examples of classes of systems satisfying this condition (e.g. see Section 4 and Table 1) Another condition **(C2)** says that a class of systems is closed under products with finite systems. We show that under **(C1)** and **(C2)**, LTL model-checking is decidable with good complexity bounds: exponential in the formula, and polynomial in the size of the input automatic presentation of the system, assuming an oracle for computing the reachability relation.

While many classes of systems satisfy **(C1)**, extending it to products (condition **(C2)**) could be problematic. Thus, we study various weakening of **(C2)** that could be used to obtain metatheorems for *fragments* of LTL. In this paper, we look at the following fragments: (1) LTL($\mathbf{F}_s, \mathbf{G}_s$) with only strict $\mathbf{F}$ and $\mathbf{G}$ operators, and (2) LTL$_{det}$ of [27]. We show that restricting to **(C2)** to closure under products with dag-like finite systems, or dropping **(C2)** altogether at the expense of a slightly worse complexity bound, decidability (and complexity) results for LTL($\mathbf{F}_s, \mathbf{G}_s$) and LTL$_{det}$ model checking could be retained. We also look at variations of these results for Presburger-definable infinite systems.

We then turn to applications, and show how our metatheorems can be used to derive (in a unified manner) known asymptotic upper bounds for LTL model-checking over some classes of systems, or produce new (or improved) complexity bounds for LTL and its fragment over other classes. Our results are summarized in Table 1.

Finally, we study the degrees of undecidability for the model checking problem and the problem of checking a relevant semantic condition over all word/tree automatic transition systems. We point out their locations in the arithmetic and analytic hierarchies.

*Organization* Definitions and notations are given in Section 2. Metatheorems are presented in Section 3. Applications are given in Section 4. Undecidability results are described in Section 5. Due to space limitations, proofs are relegated into the appendix, which can be requested from the authors.

*Related Work* The study of logical structures generated by finite-state automata and transducers can be traced far back (e.g., [15]). Since then, various models of finite-state automata and transducers have been studied, e.g., rational transducers on words (cf. [5, 29]), synchronized rational transducers on words and trees (cf. [5, 7, 12, 34]), synchronized rational transducers on infinite words and infinite trees (cf. [5, 7, 8]), and length-preserving synchronized rational transducers on finite words (cf. [2]). See [5] for a detailed comparison of their expressive power. In this paper we are concerned only with synchronized rational transducers on finite words and trees. In the case of length-preserving rational transducers on finite words, it is easy to show that LTL model checking is decidable under condition **(C1)** and **(C2)** (cf. [2]). The difficulty of extending this result to (not necessarily length-preserving) synchronized rational transducers on finite words lies in the fact that one has to deal with *genuinely* infinite execution paths, which do not visit two states twice, in the transition systems. Such paths do not exist when the length-preserving restriction is imposed on the transducers.

It is natural to ask whether our results hold in the case of the more general class of rational transducers or synchronized rational transducers on $\omega$-words (they are actually incomparable [5]). We do not know the answer to this question and leave this for future work. We also mention the paper [8], which offers a semi-algorithmic approach to handling LTL model checking over systems generated by deterministic-weak synchronized rational transducers on $\omega$-words. Finally, we mention that even though the aforementioned notions of transducers are Turing-powerful, they cannot capture all transition systems generated by higher-order pushdown systems (cf. [5, 20]).

## 2   Preliminaries

**Transition systems, reachability, and recurrent reachability** Let ACT be a finite set of *action labels*. A *transition system* over ACT is given as $\mathcal{S} = \langle S, (\to_a)_{a \in \text{ACT}} \rangle$, where $S$ is a set of *states* and each $\to_a$ is a binary relation on $S$, i.e., a subset of $S \times S$. The set $S$ is not required to be finite. We write $\to$ for the union of all transition relations $\to_a$ ($a \in \text{ACT}$) and $\to^+$ (resp. $\to^*$) to denote the transitive (resp. transitive-reflexive) closure of $\to$.

Given a transition system $\mathcal{S} = \langle S, (\to_a)_{a \in \text{ACT}} \rangle$ and a set $X \subseteq S$, by $\text{Reach}^\infty(\mathcal{S}, X)$ we denote the set of states $s \in S$ from which there exists an infinite execution path in $\mathcal{S}$ visiting $X$ infinitely often, i.e., there exists an infinite sequence $s \to^+ s_0 \to^+ s_1 \to^+ s_2 \to^+ \ldots$ so that $s_i \in X$ for all $i \geq 0$. We refer to these sets as *recurrent reachability* sets.

**Automata and transducers** We assume basic familiarity with automata on finite and $\omega$-words. Let $\Sigma$ be a finite alphabet. Given an automaton $\mathcal{A} = (Q, \delta, q_0, F)$ with states $Q$, initial state $q_0$, final states $F$ and transition function $\delta$, a run of $\mathcal{A}$ on $w = a_1 \ldots a_n$ (with $n \leq \omega$) is a function $\rho : \{0, \ldots, n\} \to Q$ with $\rho(0) = q_0$ that obeys $\delta$, i.e. $\rho(i+1) \in \delta(\rho(i), a_{i+1})$. The length $\|\rho\|$ of $\rho$ is $n$. We use abbreviations NWA and NBWA for nondeterministic (Büchi) automata.

We use *synchronized rational (letter-to-letter) transducers* [7] to define relations $P$ over $\Sigma$-words, i.e., $P \subseteq \Sigma^* \times \Sigma^*$. Such transducers are simply NWA $R$ over $\Sigma_\bot \times \Sigma_\bot$, where $\Sigma_\bot := \Sigma \cup \{\bot\}$ and $\bot \notin \Sigma$ is a padding symbol (so that the NWA could take two input words of different length). More precisely, given two words $w = a_1 \ldots a_n$ and $w' = b_1 \ldots b_m$ over the alphabet $\Sigma$, we define a word $w \otimes w'$ of length $k = \max\{n, m\}$ over alphabet $\Sigma_\bot \times \Sigma_\bot$ so that the $i$th letter of $w \otimes w'$ is $\begin{bmatrix} a'_i \\ b'_i \end{bmatrix}$, where $a'_i$ is $a_i$ for $i \leq n$, and $\bot$ for $i > n$ (and likewise $b'_i = b_i$ for $i \leq m$ and $\bot$ for $i > m$). That is, the shorter word is padded with $\bot$'s, and the $i$th letter of $w \otimes w'$ is then the pair of the $i$th letters of padded $w$ and $w'$. The binary relation "recognized" by the transducer $R$ is the set $\{(w, w') \in \Sigma^* \times \Sigma^* : w \otimes w' \in L(R)\}$. Such a relation is also called *regular*. We refer to such an automaton as a *transducer* over $\Sigma^*$, since it can be alternatively viewed as mapping words $w \in \Sigma^*$ nondeterministically into words $w'$ so that $w \otimes w'$ is accepted by $R$.

Likewise we define transducers over finite $k$-ary trees [7, 12, 34]. In the following, we recall the definition for $k = 2$. A binary tree $T = (D, \tau)$ consists of a tree domain (a finite prefix-closed subset of $\{0, 1\}^*$) and a node labeling function $\tau : D \to \Sigma$. The notation $T = T_1 \otimes T_2$ is used to refer to a tree over the labeling alphabet $\Sigma^2_\bot$ similarly to the definition of $w \otimes w'$. That is, the domain of $T$ is $D_1 \cup D_2$, and the labeling $\tau : D_1 \cup D_2 \to \Sigma^2_\bot$ is defined as $\tau(u) = (a_1, a_2)$ so that $a_i = \tau_i(u)$ if $u \in D_i$ and $\bot$ otherwise, for $i = 1, 2$. With this definition, the notion of tree transducers is defined similarly to the notion of word transducers, as a nondeterminsitic tree automaton working on $T_1 \otimes T_2$. Binary relations over trees defined that can be recognized by such transducers are called *(tree) regular*. In the sequel, we use NTA (resp. NTT) for tree automata (resp. transducers).

We shall use the notations $L(\mathcal{A})$ (or $L(R)$) for the language (or relation) accepted by automaton (or transducer) $\mathcal{A}$ (or $R$).

**Automatic presentations of transition systems** We deal with infinite transition systems that can be finitely presented by automata and transducers. A *word-automatic presentation* is $\vartheta = \langle \mathcal{A}; \{R_a\}_{a \in \text{ACT}} \rangle$ where $\mathcal{A}$ is an automaton over some finite alphabet $\Sigma$, and each $R_a$ is a transducer over $\Sigma$. This presentation generates an automatic transition system $\Lambda(\vartheta) = \langle S; \{\to_a\}_{a \in \text{ACT}} \rangle$, where $S = L(\mathcal{A})$ and $\to_a := L(R_a) \cap S$ for each $a \in \text{ACT}$. *Tree-automatic presentations* and transition systems generated by them are defined similarly except that $\mathcal{A}$ is a tree automaton and $R_a$'s are tree transducers.

Given a transition system $\langle S; \{\to_a\}_{a \in \text{ACT}} \rangle$ generated by a word or a tree-automatic presentation, each first-order (FO) formula $\varphi(x)$ with one free variable (resp. $\varphi(x, y)$ with two free variable) can effectively be converted into a word or tree automaton defining $\{s \in S \mid \varphi(s) \text{ is true}\}$ (resp. word or tree transducer defining $\{(s, s') \in S \times S : \psi(s, s') \text{ is true}\}$. This could actually be generalized to $k$ free variables [7].

We denote by $\text{wAut}_\text{P}$ and $\text{tAut}_\text{P}$ the classes of word-automatic and tree-automatic presentations, respectively. In the sequel, our metatheorems will talk about subclasses $\mathcal{C} \subseteq \text{wAut}_\text{P}$ or $\mathcal{C} \subseteq \text{tAut}_\text{P}$ satisfying certain conditions. The following several conditions will be tacitly assumed for such $\mathcal{C}$. First, it should be easy (i.e. in poly-time) to check membership in $\mathcal{C}$. This condition has a standard complexity-theoretic explanation: checking whether the input encoding of an instance to a problem is valid should be easily doable. Secondly, we do not require these classes $\mathcal{C}$ to be isomorphism-closed, i.e., there possibly exist two automatic presentations $\vartheta \in \mathcal{C}$ and $\vartheta' \in \overline{\mathcal{C}}$ generating two isomorphic transition systems $\Lambda(\vartheta)$ and $\Lambda(\vartheta')$. In fact, asserting closure under isomorphism is too strong as it is undecidable to check isomorphisms for automatic systems [7].

**LTL** The syntax of LTL over ACT is

$$\varphi, \varphi' := a \ (a \in \text{ACT}) \mid \neg \varphi \mid \varphi \vee \varphi' \mid \varphi \wedge \varphi' \mid \mathbf{X}\varphi \mid \varphi \mathbf{U} \varphi'.$$

We shall use the standard abbreviations: $\mathbf{F}\varphi$ for $\mathit{true}\mathbf{U}\varphi$, $\mathbf{G}\varphi$ for $\neg\mathbf{F}\neg\varphi$, and $\mathbf{F}_s$ and $\mathbf{G}_s$ for their strict versions: $\mathbf{F}_s\varphi = \mathbf{X}\mathbf{F}\varphi$ and $\mathbf{G}_s\varphi = \neg\mathbf{F}_s\neg\varphi$.

Given an $\omega$-word $w \in \mathrm{ACT}^\omega$, we define the satisfaction relation $w \models \varphi$ in the standard way. We write $[\![\varphi]\!]$ for the set of all $w \in \mathrm{ACT}^\omega$ such that $w \models \varphi$.

It is well-known [35] that there exists an exponential-time algorithm which, given an LTL formula $\varphi$, computes an NBWA $\mathcal{A}_\varphi$ satisfying $L(\mathcal{A}_\varphi) = [\![\varphi]\!]$.

Given a transition system $\mathcal{S} = \langle S, (\to_a)_{a \in \mathrm{ACT}} \rangle$ and a word $u = a_0 a_1 a_2 \ldots \in \mathrm{ACT}^\omega$, we say that $w_0 \in S$ *realizes* $u$ if there is a sequence of $w_0, w_1, w_2, \ldots$ of elements of $S$ so that $w_i \xrightarrow{a_i} w_{i+1}$ for all $i \geq 0$. We then define the semantics of LTL formulae in the standard way: $(\mathcal{S}, w_0) \models \varphi$ iff every $\omega$-word $u \in \mathrm{ACT}^\omega$ realized by $w_0$ satisfies $\varphi$. We write $[\![\varphi]\!]_\mathcal{S}^\forall$ for the set of all $w_0 \in S$ such that $(\mathcal{S}, w_0) \models \varphi$ (where $\forall$ means that every path starting in $w_0$ satisfies $\varphi$). We also write $[\![\varphi]\!]_\mathcal{S}^\exists$ for the complement of the set $[\![\neg\varphi]\!]_\mathcal{S}^\forall$, i.e., for the set of $w_0$ that realizes at least one path satisfying $\varphi$.

## 3  Metatheorems for LTL and its fragments

Since LTL formulae are translated into Büchi automata, a starting point for us is a known metatheorem that gives a semantic condition which implies bounds (and structural properties) for recurrent reachability sets. We now define condition on a class $\mathcal{C}$ of presentations in $\mathrm{wAut}_\mathrm{P}$ (resp. in $\mathrm{tAut}_\mathrm{P}$):

> **(C1)** There exists an algorithm $A_\mathcal{C}$ which, given an input presentation $\vartheta = \langle \mathcal{A}; \{R_a\}_{a \in \mathrm{ACT}} \rangle \in \mathcal{C}$ of the automatic transition system $\Lambda(\vartheta) = \langle S; \{\to_a\}_{a \in \mathrm{ACT}} \rangle$, computes an NWT (resp. NTT) $R^+$ recognizing the transitive closure relation $\to^+ = \left( \bigcup_{a \in \mathrm{ACT}} \to_a \right)^+$.

Intuitively, **(C1)** asserts that the transitive closure relations of systems $\Lambda(\vartheta)$ with $\vartheta \in \mathcal{C}$ are effectively regular. We denote the running time of $A_\mathcal{C}$ to be $t_{A_\mathcal{C}}$. The following results state that under **(C1)**, recurrent reachability sets can be computed with polynomial-time overhead.

**Theorem 1 ([34]).** *Fix any class $\mathcal{C} \subseteq \mathrm{wAut}_\mathrm{P}$ (resp. $\mathcal{C} \subseteq \mathrm{tAut}_\mathrm{P}$) satisfying (C1). Given an automatic presentation $\vartheta \in \mathcal{C}$ and an NWA (resp. NTA) $\mathcal{A}_0$, the set $\mathrm{Reach}^\infty(\Lambda(\vartheta), L(\mathcal{A}_0))$ is regular, for which an NWA (resp. NTA) is computable in time polynomial in $\|\vartheta\| + \|\mathcal{A}_0\| + t_{A_\mathcal{C}}(|\vartheta|)$. In particular, if $A_\mathcal{C}$ runs in poly-time, then an NWA (resp. NTA) for $\mathrm{Reach}^\infty(\Lambda(\vartheta), L(\mathcal{A}_0))$ is poly-time computable.*

### 3.1  A metatheorem for LTL

We now adapt Theorem 1 to produce a metatheorem for LTL. Consider a finite system $\mathcal{F} = \langle Q = \{q_0, \ldots, q_n\}, \delta \rangle$, with $\delta : Q \times \mathrm{ACT} \to Q$. Given a presentation $\vartheta \in \mathrm{wAut}_\mathrm{P}$ of the system $\Lambda(\vartheta) = \langle S \subseteq \Sigma^*, \{\to_a\}_{a \in \mathrm{ACT}} \rangle$, we define $\mathcal{F} \cdot \Lambda(\vartheta)$ to be the automatic transition system $\langle S'; \{\to'_a\}_{a \in \mathrm{ACT}} \rangle$ as follows:

- $S' := QS := \{qs : q \in Q, s \in S\}$; it is a regular language over $S \cup Q$.

– $qw \to'_a q'w'$ iff $q' \in \delta(q,a)$ and $w \to_a w'$.

It is easy to give an automatic presentation $\vartheta'$ of $\mathcal{F} \cdot \Lambda(\vartheta)$ and show that $\vartheta'$ is poly-time computable. For presentations $\vartheta \in \text{tAut}_\text{P}$, we could define $\mathcal{F} \cdot \Lambda(\vartheta)$ in a similar way (e.g. by defining the domain to be $Q(S) = \{q(T) : q \in Q, T \in S\}$ where $q(T)$ is the tree obtained by attaching $q$ to $T$ as a root).

We now define another condition **(C2)** stating that the class $\mathcal{C}$ is closed under products with finite systems:

> **(C2)** if $\vartheta \in \mathcal{C}$ and $\mathcal{F}$ is a finite system then $\mathcal{F} \cdot \Lambda(\vartheta) \in \mathcal{C}$.

The following theorem is now almost immediate from Theorem 1 and the standard translation from LTL into Büchi automata. Intuitively, it says that for automatic presentations satisfying both **(C1)** and **(C2)**, for LTL model-checking the overhead (compared to $t_{A_\mathcal{C}}$) is polynomial in the automatic presentation and exponential in the LTL formula. In particular, if $t_{A_\mathcal{C}}$ is polynomial itself, then LTL model-checking is polynomial in the size of the representation of the system and exponential in the size of the formula.

**Theorem 2.** *Fix any set $\mathcal{C} \subseteq \text{wAut}_\text{P}$ (resp. $\mathcal{C} \subseteq \text{tAut}_\text{P}$) satisfying both **(C1)** and **(C2)**. Given $\vartheta \in \mathcal{C}$ and an LTL formula $\varphi$, the set $[\![\neg\varphi]\!]^\exists_{\Lambda(\vartheta)}$ is regular, for which an automaton is computable in time polynomial in $\|\vartheta\| + \|\mathcal{A}\| + t_{A_\mathcal{C}}(2^{O(\|\varphi\|)} \times \|\vartheta\|)$. Thus, checking whether $(\Lambda(\vartheta), v_0) \models \varphi$ can be done in time polynomial in $\|\vartheta\| + \|v_0\| + t_{A_\mathcal{C}}(2^{O(\|\varphi\|)} \times \|\vartheta\|)$.*

There are many examples of classes of automatic structures of interest in verification that satisfy condition **(C1)** (see, e.g., [34] for a list). So it is natural to ask whether having condition **(C1)** for a class of automatic presentations $\mathcal{C}$ implies having it for products of structures in that class with finite systems. While we shall see some examples of classes where this happens (e.g., pushdown systems), in general such an extension is impossible even in very simple cases, e.g., for single structures, as the result below shows.

**Proposition 3.** *There exist an automatic presentation $\vartheta$ satisfying **(C1)** and a finite system $\mathcal{F}$ so that in $\mathcal{F} \cdot \Lambda(\vartheta)$ the reachability relation is not regular (in fact, not even recursive).*

So the applicability of Theorem 2 in full generality may be rather limited. We thus look at cases when conditions weaker than **(C2)** will allow us to conclude the decidability of model-checking. They will not apply to full LTL, but they will apply to some of its well-studied fragments. The distinguishing feature of these fragments is that formulae in them can be translated into special types of Büchi automata, whose graph structures are rather nice (essentially, almost DAGs). We next look at such cases.

### 3.2 Metatheorems for LTL$_\text{det}$

We first recall the definition of the fragment LTL$_\text{det}$ of LTL [27].

$$\varphi, \varphi' := p \mid \mathbf{X}\varphi \mid \varphi \wedge \varphi' \mid (p \wedge \varphi) \vee (\neg p \wedge \varphi') \mid$$
$$(p \wedge \varphi)\mathbf{U}(\neg p \wedge \varphi') \mid (p \wedge \varphi)\mathbf{W}(\neg p \wedge \varphi').$$

Here $p$ is a boolean combination of ACT, and $\varphi \mathbf{W} \varphi'$ is interpreted as the formula $\mathbf{G}\varphi \vee (\varphi \mathbf{U} \varphi')$, i.e., the "weak until" operator.

Formulae in this fragment can be translated into a special kind of automata called 1-weak NBWAs. Formally, a *1-weak NBWA* $\mathcal{A} = (\Sigma, Q, \delta, q_0, F)$ is an NBWA with a partial order $\preceq \subseteq Q \times Q$ such that $q' \in \delta(q, a)$ implies $q \preceq q'$. Intuitively, the partial order ensures that once $\mathcal{A}$ leaves a state $q$, it will never be able to come back to $q$. In other words, graph-theoretically $\mathcal{A}$ looks like a dag possibly with self-loops.

It was shown in [27] that there exists a poly-time algorithm which, given an $LTL_{det}$ formula $\varphi$, computes a 1-weak NBWA $\mathcal{A}_{\neg\varphi}$ such that $L(\mathcal{A}_{\neg\varphi}) = [\![\neg\varphi]\!]$. (The running time was not explicitly mentioned in [27], but one can easily check that it is polynomial).

We now weaken the condition **(C2)** to the following:

> **(C2')** if $\vartheta \in \mathcal{C}$ and $\mathcal{F}$ is a finite system that is 1-weak then $\mathcal{F} \cdot \Lambda(\vartheta) \in \mathcal{C}$.

Combining Theorem 1 with the translation of [27], we may proceed as in the proof of Theorem 2 and obtain the following theorem.

**Theorem 4.** *Fix any set $\mathcal{C} \subseteq \text{wAut}_\text{P}$ (resp. $\mathcal{C} \subseteq \text{tAut}_\text{P}$) satisfying (C1) and (C2'). Given $\vartheta \in \mathcal{C}$ and an $LTL_{det}$ formula $\varphi$, the set $[\![\neg\varphi]\!]^\exists_{\Lambda(\vartheta)}$ is regular, for which an automaton is computable in time polynomial in $\|\vartheta\| + \|\mathcal{A}\| + t_{A_\mathcal{C}}(\|\varphi\| \times \|\vartheta\|)$. Thus, checking whether $(\Lambda(\vartheta), v_0) \models \varphi$ can be done in time polynomial in $\|\vartheta\| + \|v_0\| + t_{A_\mathcal{C}}(\|\varphi\| \times \|\vartheta\|)$.*

We now show that decidability can still be obtained without assuming condition **(C2')** but by slightly strengthening condition **(C1)**. Namely, we use a condition stating that the transitive closure can be computed not only for $\to$ but also for all unions of $\to_a$'s:

> **(C1')** there exists an algorithm $A_\mathcal{C}$ which, given an input presentation $\vartheta = \langle \mathcal{A}_\delta; \{R_a\}_{a \in \text{ACT}} \rangle \in \mathcal{C}$ of the automatic transition system $\Lambda(\vartheta) = \langle S; \{\to_a\}_{a \in \text{ACT}} \rangle$ and each subset $\text{ACT}' \subseteq \text{ACT}$, computes an NWT (resp. NTT) $R^+_{\text{ACT}'}$ recognizing the transitive closure relation $\left(\bigcup_{a \in \text{ACT}'} \to_a\right)^+$.

In practice, **(C1')** is not much stronger than **(C1)**; all our examples in the next section which satisfy **(C1)** also satisfy **(C1')**. In this case, $LTL_{det}$ model-checking can be done in PSPACE assuming an oracle for $t_{A_\mathcal{C}}$; its running time is only exponential in the the size of the formula.

**Theorem 5.** *Fix any set $\mathcal{C} \subseteq \text{wAut}_\text{P}$ (resp. $\mathcal{C} \subseteq \text{tAut}_\text{P}$) satisfying (C1'). Given a presentation $\vartheta \in \mathcal{C}$, and an $LTL_{det}$ formula $\varphi$, checking whether $(\Lambda(\vartheta), v_0) \models \varphi$ can be done in time polynomial in $\|\vartheta\|$, $\|v_0\|$, $t_{A_\mathcal{C}}(\|\vartheta\|)$, and exponential in $\|\varphi\|$. Whenever $\mathcal{C} \subseteq \text{wAut}_\text{P}$, the space consumed by the algorithm is polynomial in $\|\vartheta\|$, $\|v_0\|$, $t_{A_\mathcal{C}}(\|\vartheta\|)$, and $\|\varphi\|$.*

### 3.3  Metatheorems for LTL($\mathbf{F_s}, \mathbf{G_s}$)

Recall that in LTL($\mathbf{F_s}, \mathbf{G_s}$) we use operators $\mathbf{F_s}$ and $\mathbf{G_s}$ rather than $\mathbf{U}$ and $\mathbf{X}$. It turns out that our conditions **(C1)** and **(C2)** imply bounds on LTL($\mathbf{F_s}, \mathbf{G_s}$) model-checking. We start with the following.

**Theorem 6.** *Fix any set $\mathcal{C} \subseteq \text{WAUT}_P$ (resp. $\mathcal{C} \subseteq \text{TAUT}_P$) satisfying **(C1)** and **(C2)**. Given a presentation $\vartheta \in \mathcal{C}$ and an LTL($\mathbf{F}_s, \mathbf{G}_s$) formula $\varphi$, checking whether $(\Lambda(\vartheta), v_0) \models \varphi$ can be done in coNP assuming an oracle for $t_{A_\mathcal{C}}$. More precisely, checking whether $(\Lambda(\vartheta), v_0) \not\models \varphi$ can be done in nondeterministic time polynomial in $\|\vartheta\| + \|v_0\| + t_{A_\mathcal{C}}(\|\varphi\| \times \|\vartheta\|)$.*

The proof of this result is based on a translated into 1-weak NBWAs extended with fairness constraints, which are conjunctions of formulae $\mathbf{G_s F_s}p$, where $p$ is a disjunction over action labels in ACT [31]. We have to extend translation results from [31] to obtain more precise information about the structure of the automata, and then use it to prove the result along the lines of the proof in the previous section. Details are in the appendix.

Our second metatheorem for LTL($\mathbf{F_s}, \mathbf{G_s}$) uses only condition **(C1)** and produces slightly higher, but still acceptable, complexity bounds.

**Theorem 7.** *Fix any set $\mathcal{C} \subseteq \text{WAUT}_P$ (resp. $\mathcal{C} \subseteq \text{TAUT}_P$) satisfying **(C1)**. Given a presentation $\vartheta \in \mathcal{C}$, and an LTL($\mathbf{F}_s, \mathbf{G}_s$) formula $\varphi$, checking whether $(\Lambda(\vartheta), v_0) \models \varphi$ can be done in time polynomial in $\|\vartheta\|$, $\|v_0\|$, $t_{A_\mathcal{C}}(\|\vartheta\|)$, and exponential in $\|\varphi\|$.*

### 3.4  A metatheorem for Presburger-definable systems

In this subsection, we will make an extra assumption that the input presentations can be given by *existential* Presburger formulas. More precisely, we consider presentations of the form $\vartheta = \langle \varphi(\overline{x}); \{\varphi_a(\overline{x}, \overline{y})\}_{a \in \text{ACT}} \rangle$, where $\overline{x}$ and $\overline{y}$ are $k$-tuples of variables for some $k \in \mathbb{Z}_{>0}$ and $\varphi$'s some existential Presburger formulas. Such a presentation gives rise to the system $\Lambda(\vartheta) = \langle S; \{\rightarrow_a\}_{a \in \text{ACT}} \rangle$, where $S = \{\overline{a} \in \mathbb{N}^k : (\mathbb{N}, +) \models \varphi(\overline{a})\}$ and $\rightarrow_a = \{(\overline{a}, \overline{b}) \in \mathbb{N}^{2k} : (\mathbb{N}, +) \models \varphi_a(\overline{a}, \overline{b})\}$. Let PRESAUT$_P$ denote the set of all such presentations. Automatic presentations for PRESAUT$_P$ could be given (cf. [7]).

For sets $\mathcal{C} \subseteq \text{PRESAUT}_P$ (which, as before, need not be isomorphism-closed), we define a new semantic condition, which is essentially an adaption of **(C1')** to the class of Presburger-definable systems:

> **(C3)** *there exists an algorithm $A_\mathcal{C}$ which, given an input presentation $\vartheta = \langle \varphi; \{\varphi_a\}_{a \in \text{ACT}} \rangle \in \mathcal{C}$ of the system $\Lambda(\vartheta) = \langle S; \{\rightarrow_a\}_{a \in \text{ACT}} \rangle$ and a subset $\text{ACT}' \subseteq \text{ACT}$, computes an existential Presburger formula $R^+(\overline{x}, \overline{y})$ which defines the transitive closure relation $\left(\bigcup_{a \in \text{ACT}'} \rightarrow_a\right)^+$.*

We denote by $t_{A_\mathcal{C}}$ the running time of $A_\mathcal{C}$ in **(C3)**. In addition, we require that the class $\mathcal{C}$ satisfy the following *monotonicity* condition: for every $\mathcal{S} \in \mathcal{C}$ every $\overline{a}, \overline{b} \in \mathbb{N}^k$ satisfying $\overline{a} \preceq \overline{b}$ (i.e. inequality holds component-wise), if $\overline{a} \to_a \overline{a} + \overline{\delta}$ for some $\overline{\delta} \in \mathbb{Z}^k$ and $a \in \text{ACT}$, then $\overline{b} \to_a \overline{b} + \overline{\delta}$. This is a strong condition, but is still satisfied by any subclass of Petri nets.

**Theorem 8.** *Fix any monotone $\mathcal{C} \subseteq \text{PRESAUT}_P$ satisfying (C3). Given $\vartheta \in \mathcal{C}$, $v_0 \in \mathbb{N}^k$ represented in binary, and an $LTL(\mathbf{F}_s, \mathbf{G}_s)$ or $LTL_{det}$ formula $\psi$, checking whether $(\Lambda(\vartheta), v_0) \not\models \psi$ can be done in nondeterministic time polynomial in $\|\vartheta\|$, $\|v_0\|$, $t_{A_\mathcal{C}}(\|\vartheta\|)$ and $\|\varphi\|$*

## 4  Applications

In this section we apply our metatheorems from the previous section to obtain decidability and complexity results for LTL, LTL($\mathbf{F}_s, \mathbf{G}_s$), and LTL$_{\text{det}}$ model checking over specific classes of infinite systems. In some cases, we re-derive known results with asymptotically the same complexity bounds; in other cases we obtain new results. Our results are summarized in Table. 1.

**Pushdown systems (PDS).** A *pushdown system* [9] is a tuple $\mathcal{P} = (\text{ACT}, \Gamma, Q, \Delta)$ where $\Gamma$ is a *stack alphabet*, $Q$ a set of states, and $\Delta$ is a finite subset of $Q \times \Gamma \times \text{ACT} \times Q \times \Gamma^*$. Let $\Delta_a$ be the set of tuples (called "rules") in $\Delta$ of the form $(q, \sigma, a, q', w)$. Let $\Lambda(\mathcal{P})$ be the transition system $\langle Q \times \Gamma^*; \{\to_a\}_{a \in \text{ACT}}\rangle$, where $\to_a := \{((q, w\sigma), (q', ww')) : (q, \sigma, a, q', w') \in \Delta_a\}$. It is straightforward (and in poly-time) to give a word-automatic presentation of $\mathcal{P}$ (cf. [34]), and show that the class PDS of such presentations satisfy **(C2)**. Furthermore, it is known [10, 34] that PDS satisfies **(C1)** with polynomial running time.

Combined with our results in the previous section, it follows that model checking LTL, LTL($\mathbf{F}_s, \mathbf{G}_s$), and LTL$_{\text{det}}$ are respectively in EXPTIME, coNP, and PTIME. It also follows that all these problems are in PTIME when fixing the formula. It was known (cf. [9]) that the complexity of model checking LTL over pushdown systems is EXPTIME-complete, and is PTIME for a fixed formula. On the other hand, the results for LTL($\mathbf{F}_s, \mathbf{G}_s$) and LTL$_{\text{det}}$ are new (in the case of LTL($\mathbf{F}_s, \mathbf{G}_s$) coNP-hardness can be derived from [32]).

**Prefix-recognizable systems (Pref-RS).** A *prefix-recognizable system (with states)* $\mathcal{P}$ is a tuple $\langle \text{ACT}, \Gamma, Q, \Delta\rangle$ where ACT, $\Gamma$, and $Q$ are defined as in pushdown systems, whereas $\Delta$ is a set of rules of the form $((q, U, V), a, (q', V'))$, where $q, q' \in Q$; $a \in \text{ACT}$; and $U, V$, and $V'$ are regular languages over $\Gamma$ given by NWAs. Let $\Lambda(\mathcal{P})$ be the transition system $S = \langle Q \times \Gamma^*; \{\to_a\}_{a \in \text{ACT}}\rangle$, where $\to_a$ is the set of tuples $((q, uv), (q', uv')) \in S \times S$ such that, for some $((q, U, V), a, (q', V')) \in \Delta_a$, we have $u \in U$, $v \in V$, and $v' \in V'$. It is straightforward (and in poly-time) to give a word-automatic presentation for $\mathcal{P}$.

Using Theorem 2, we can also rederive the known EXPTIME upper bound [22] for LTL model checking over prefix-recognizable systems (details are in the appendix). Furthermore, EXPTIME-hardness for model checking a fixed $\text{LTL}_{\text{det}}$ and $\text{LTL}(\mathbf{F}_s, \mathbf{G}_s)$ formula is obtained by reducing from the unreachability problem for prefix-recognizable systems, which has recently been proven to be EXPTIME-complete [18].

**Concurrent pushdown systems (CPDS).** A concurrent pushdown system (cf. [30]) is a tuple $\mathcal{P} = (\text{ACT}, \Gamma, Q, \Delta^0, \ldots, \Delta^N)$, where each $\mathcal{P}_i = (\text{ACT}, \Gamma, Q, \Delta^i)$ is a pushdown system. Suppose that $\Lambda(\mathcal{P}_i) = \langle Q \times \Gamma^*; \{\rightarrow_{i,a}\}_{a \in \Gamma}\rangle$. Then the transition system $\Lambda(\mathcal{P})$ generated by $\mathcal{P}$ is $\langle Q \times (\Gamma^*)^N; \{\rightarrow_a\}_{a \in \Gamma}\rangle$, where $\rightarrow_a := \bigcup_{i=0}^{N} \rightarrow_{i,a}$. Although concurrent pushdown systems are well-known to be Turing-powerful (so checking safety and liveness is undecidable), [23, 30] have shown that reachability is NP-complete if we consider runs of $\mathcal{P}$ with a bounded number $k$ of "context-switches" ($k$ part of the input). Intuitively, a context of $\mathcal{P}$ is an uninterrupted sequence of actions performed by exactly one "thread" $\mathcal{P}_i$. A context-switch occurs when $\mathcal{P}$ interrupts the execution of a thread $\mathcal{P}_i$ and resumes by executing a (possibly different) thread $\mathcal{P}_j$. The context-bounded reachability for $\mathcal{P}$ is simply the problem of reachability restricted to executions of $\mathcal{P}$ with exactly $k$ context-switches for any given input $k$. One could similarly define the context-bounded LTL model checking problem for concurrent pushdown systems by restricting the executions of $\mathcal{P}$ to those with exactly $k$ context-switches for any given $k$.

Using the results of [23, 30] and our metatheorems, we can show that context-bounded model checking LTL, $\text{LTL}(\mathbf{F}_s, \mathbf{G}_s)$, and $\text{LTL}_{\text{det}}$ over concurrent pushdown systems are respectively EXPTIME-complete, coNP-complete, and coNP-complete. If the formula is fixed, they are all coNP-complete.

**Discrete timed counter systems (RCM and d-RCM).** Although verifying safety and liveness for general counter machines is undecidable, it is known that these problems are decidable (cf. [14, 21]) when all the counters but one are *reversal-bounded* (only executions with a fixed number of reversals are considered). We denote by RCM the class of such machines. The LTL model checking problem for RCM is also known to be decidable [14], but no complexity analysis was given for their algorithm. Furthermore, it was left as an open question whether the same result holds for such machines extended with discrete clocks (in the sense of [3]), for which reachability is known to be decidable [13]. We write d-RCM for the class of such machines.

We answer this open question positively and give upper bounds for the cases with and without discrete clocks. Using our metatheorems in combination with a slightly refined version of the algorithms for computing binary reachability relations in [13, 21], we can give an algorithm for model checking LTL (resp. $\text{LTL}_{\text{det}}$ and $\text{LTL}(\mathbf{F}_s, \mathbf{G}_s)$) over RCM that runs in time exponential in the size of the machine and double exponential (resp. exponential) in the size of the LTL (resp. $\text{LTL}_{\text{det}}$ and $\text{LTL}(\mathbf{F}_s, \mathbf{G}_s)$) formula. Details of the construction and the analysis are in the appendix.

For d-RCM, we have exactly the same upper bound complexity except that the algorithms run double exponential in the number of clocks.

**Communication-free Petri nets (BPP).** *Communication-free nets* (a.k.a. BPPs) [16, 28] are Petri nets where each transition has exactly one incoming arc (and, hence, "communication-free"). The LTL model checking over BPPs is known to be EXPSPACE-complete when only infinite traces are considered (cf. [28]). When finite traces are also considered, the problem is still decidable but no primitive recursive upper bound is known [28], since reachability for Petri nets could be reduced to this problem.

In contrast, we could show that $\text{LTL}_{\text{det}}$ and $\text{LTL}(\mathbf{F}_s, \mathbf{G}_s)$ model checking for BPPs is coNP-complete even when finite traces are considered. In fact, it is known that the transitive closure relations for BPPs are semi-linear [16]. Furthermore, one can then adapt the proof of [36, Theorem 4] to show that there exists a poly-time algorithm computing an existential Presburger formula defining the transitive closure relation for a given BPP. Since any subclass of Petri nets is monotone, Theorem 8 (which also holds when finite traces are considered) implies that $\text{LTL}(\mathbf{F}_s, \mathbf{G}_s)$ and $\text{LTL}_{\text{det}}$ model checking over BPPs are in coNP.

Furthermore, a matching lower bound could be easily given for a fixed formula in $\text{LTL}(\mathbf{F}_s, \mathbf{G}_s)$ and $\text{LTL}_{\text{det}}$ by reducing from the non-reachability problem for BPPs, which is coNP-complete [16].

**Weakly extended PA processes (wPA).** PA (cf. [26, 28, 31]) is a well-known process algebra allowing sequential and parallel compositions, but no communication. It is a common generalization of BPP (with unary representation for numbers) and the class of pushdown systems with one state (a.k.a. BPA). It is known (cf. [28, 31]) that LTL model checking over PA is undecidable. It is also known that decidability could be retained when restricting to $\text{LTL}(\mathbf{F}_s, \mathbf{G}_s)$ and $\text{LTL}_{\text{det}}$ [31]. However, no upper bound to these problems are known.

We can use Theorem 5 and Theorem 7 in combination with the encoding of PA and their binary reachability relations as tree-automatic systems (cf. [26, 34]) to give an exponential time upper bound for these problems. They are coNP-hard, which can be shown by a reduction from non-reachability problem for BPP [16]. The upper bound also holds when we consider *weakly extended PA* (wPA) [31], which are simply PA extended with weak finite control (i.e. 1-weak NBWA).

**Weakly extended ground-tree rewrite systems (wGTRS).** A *ground tree rewrite system* (GTRS) (cf. [25, 34]) over $\Sigma$-labeled trees is a finite set $\Delta$ of "rules" of the form $(t, a, t')$ where $t, t' \in \text{Tree}(\Sigma)$ and $a \in \text{ACT}$. For a tree $T$ and a node $u$ in it, let $T_u$ be the subtree of $T$ rooted at $u$. For a given $t \in \text{Tree}(\Sigma)$, we write $T[t/u]$ for the tree obtained from $T$ by replacing the subtree $T_u$ by $t$. The GTRS $\Delta$ generates the transition system $\Lambda(\Delta) = \langle \text{Tree}(\Sigma); \{\rightarrow_a\}_{a \in \text{ACT}} \rangle$ where $T \rightarrow_a T'$ iff there exists a node $u$ in $T$ and a rule $(t, a, t') \in \Delta$ such that $T_u = t$ and $T' = T[t'/u]$. One could easily conclude that LTL model checking over GTRS is undecidable, using results of [25, 31].

On the other hand, our results imply that decidability is retained when we restrict to $\text{LTL}(\mathbf{F}_s, \mathbf{G}_s)$ or $\text{LTL}_{\text{det}}$. This follows from the fact that **(C1')** is satisfied by the class of automatic presentations of GTRSs (cf. [12, 34]). Therefore, we obtain exponential-time algorithms for model checking LTL$_{\text{det}}$ and LTL($\mathbf{F}_s, \mathbf{G}_s$) over GTRS, whose complexity becomes polynomial when the formula is fixed. We could also show that these problems are coNP-hard for non-fixed formulas.

One can also extend these results to GTRSs with weak finite control, as we did for PA-processes. Details are in the appendix.

|         | LTL       |           | LTL($\mathbf{F}_s, \mathbf{G}_s$) |           | LTL$_{\text{det}}$ |           |
|---------|-----------|-----------|-----------|-----------|-----------|-----------|
|         | Comb.     | Data      | Comb.     | Data      | Comb.     | Data      |
| PDS     | EXP       | in P      | **coNP**  | in P      | **in P**  | in P      |
| Pref-RS | EXP       | EXP       | EXP       | EXP       | EXP       | EXP       |
| CPDS    | **EXP**   | **coNP**  | **coNP**  | **coNP**  | **coNP**  | **coNP**  |
| BPP     | ×         | ×         | **coNP**  | **coNP**  | **coNP**  | **coNP**  |
| (w)PA   | × (ud)    | × (ud)    | **in EXP** coNP-h | **in EXP** coNP-h | **in EXP** coNP-h | **in EXP** coNP-h |
| GTRS    | × (ud)    | × (ud)    | **in EXP** coNP-h | **in P**  | **in EXP** coNP-h | **in P**  |
| wGTRS   | × (ud)    | × (ud)    | **in EXP coNP-h** | **in EXP coNP-h** | **in EXP coNP-h** | **in EXP coNP-h** |
| RCM     | **in 2-EXP** | **in EXP** | **in EXP** | **in EXP** | **in EXP** | **in EXP** |
| d-RCM   | **in 2-EXP** |||||||

**Table 1.** A summary of combined and data complexity that we obtain. Here, × (resp. ud) means that the result cannot be obtained using our metatheorems (resp. undecidable). Whenever written in bold, the results are new. Also, coNP-h means coNP-hard.

## 5  How hard are these problems in general?

Relevant to condition **(C1)** is the problem of checking whether the transitive closure relation of a given automatic presentation is regular, and the problem of checking whether a given transducer $R'$ represents the transitive closure relation of another one $R$ (over the same domain). We shall point out the degrees of undecidability of such problems in the arithmetic hierarchy. We shall then point out the degrees of undecidability of the model checking problems in the general case (i.e. over all word/tree automatic presentations), and compare this with the length-preserving case. We start with the problems related to "computing" transitive closure relations.

**Theorem 9.**  – *Given two nondeterministic word transducers $R$ and $R'$, checking whether $R'$ is the transitive closure of $R$ is $\Pi_2^0$-complete.*
 – *Given a nondeterministic word transducers $R$, checking whether its transitive closure is regular is in $\Sigma_3^0$ and $\Pi_2^0$-hard.*

We now address the degrees of undecidability for checking recurrent reachability and model checking LTL, LTL($\mathbf{F}_s, \mathbf{G}_s$), and LTL$_{\text{det}}$ over automatic transition systems. Unlike the problem of reachability which can be shown to be $\Sigma_1^0$-complete (cf. [7, 29]), checking liveness is highly undecidable:

**Theorem 10.** *Recurrent reachability for both word and tree automatic transition systems is $\Sigma_1^1$-complete, and model-checking LTL, $LTL(\mathbf{F}_s, \mathbf{G}_s)$, and $LTL_{det}$ for them are all $\Pi_1^1$-complete.*

In fact, Theorem 10 could be shown to also hold when the general class of rational transducers is used (instead of synchronized rational).

Finally, many examples in the "regular model checking" literature (cf. [2]) deal with the subcase of length-preserving synchronized rational transducers (i.e., $w \rightarrow_a w'$ would imply $|w| = |w'|$). In this case, the LTL (resp. recurrent reachability) model checking problem is usually defined with respect to a regular set $Init$ of initial states with either "existential" (resp. "universal") semantics in the following sense: there exists $w \in Init$ such that (resp. all $w \in Init$ satisfies) $(\Lambda(\vartheta), w) \models \varphi$. [Note: when $Init$ is finite, the model checking problems become decidable since then the set of reachable states from Init is finite.] In contrast to Theorem 10, we have the following proposition.

**Proposition 11.** *For automatic transition systems with length-preserving transducers, global recurrent reachability and LTL model checking are all*

- *$\Sigma_1^0$-complete (when existential semantics is considered);*
- *$\Pi_1^0$-complete (when universal semantics is considered).*

This result confirms the intuition that checking liveness is much easier when considering length-preserving transducers.

## References

1. P. A. Abdulla, and B. Jonsson. Verifying programs with unreliable channels. *Inf. Comput.*, 127(2):91–101, 1996.
2. P. A. Abdulla, B. Jonsson, M. Nilsson, and M. Saksena. A survey of regular model checking. In *CONCUR'04*, pages 35–48.
3. R. Alur and D. Dill. A theory of timed automata. *TCS*, 126 (1994), 183–235.
4. C. Baier, N. Bertrand, and Ph. Schnoebelen. On Computing Fixpoints in Well-Structured Regular Model Checking, with Applications to Lossy Channel Systems. In *LPAR'06*, pages 347–361.
5. V. Barany. Automatic Presentations of Infinite Structures. PhD Thesis, RWTH Aachen. 2007.
6. S. Bardin, A. Finkel, J. Leroux, L. Petrucci. FAST: acceleration from theory to practice. *STTT* 10(5): 401–424 (2008)
7. A. Blumensath and E. Grädel. Finite presentations of infinite structures: automata and interpretations. *Theory Comput. Syst.* 37(6): 641–674 (2004)
8. A. Bouajjani, A. Legay, and P. Wolper. A Framework to Handle Linear Temporal Properties in ($\omega$)Regular Model Checking *CoRR* abs/0901.4080: (2009)
9. O. Burkart, D. Caucal, F. Moller and B. Steffen. Verification on infinite structures. In *Handbook of Process Algebra*, Elsevier, 1999.
10. D. Caucal. On the regular structure of prefix rewriting. *TCS*, 106 (1992), 61–86.
11. H. Comon and Y. Jurski. Multiple counters automata, safety analysis and Presburger arithmetic. In *CAV'98*, pages 268–279.

12. H. Comon et al. *Tree Automata: Techniques and Applications*. Available on: http://www.grappa.univ-lille3.fr/tata, 2007.
13. Z. Dang, O. Ibarra, T. Bultan, R. Kemmerer, J. Su. Binary reachability analysis of discrete pushdown timed automata. In *CAV 2000*, pages 69–84.
14. Z. Dang, O. Ibarra, P.S. Pietro. Liveness verification of reversal-bounded multi-counter machines with a free counter. In *FSTTCS'01*, pages 132–143.
15. C. Elgot and J. Mezei. On relations defined by generalized finite automata. *IBM J. Res. Develop.* 9 (1965), 47–68.
16. J. Esparza. Petri Nets, Commutative Context-Free Grammars, and Basic Parallel Processes. *Fundam. Inform.* 31(1): 13-25 (1997)
17. A. Finkel and Ph. Schnoebelen. Well-structured transition systems everywhere! Theor. Comput. Sci. 256(1-2):63–92 (2001)
18. S. Göller. Reachability on prefix-recognizable graphs. *Inf. Process. Lett.* 108(2): 71-74 (2008)
19. M. Grohe. Logic, graphs, and algorithms. In *Logic and Automata - History and Perspectives*, pages 357–422, Amsterdam University Press, 2007.
20. M. Hague and C.-H.L. Ong. Symbolic Backwards-Reachability Analysis for Higher-Order Pushdown Systems. *LMCS* 4(4): (2008)
21. O. Ibarra, J. Su, Z. Dang, T. Bultan, R. Kemmerer. Counter machines and verification problems. *Theor. Comput. Sci.* 289 (2002), 165–189.
22. O. Kupferman, N. Piterman, M. Vardi. Model checking linear properties of prefix-recognizable systems. In *CAV 2002*, pages 371–385.
23. A. Lal, T. Touili, N. Kidd and T. Reps. Interprocedural Analysis of Concurrent Programs Under a Context Bound. In *TACAS'08*, pages 282–298.
24. J. Leroux and G. Sutre. Flat Counter Automata Almost Everywhere! In *ATVA '05*, pages 489–503.
25. C. Löding. Infinite Graphs Generated by Tree Rewriting, PhD thesis, RWTH Aachen, 2003.
26. D. Lugiez and P. Schnoebelen. Decidable first-order transition logics for PA-processes. *Inf. Comput.*, 203(1):75–113, 2005.
27. M. Maidl. The Common Fragment of CTL and LTL. In *FOCS'00*, pages 643–652.
28. R. Mayr. Decidability and Complexity of Model Checking Problems for Infinite-State Systems. PhD thesis, TU-Munich, 1998.
29. C. Morvan. On rational graphs. In *FOSSACS'00*, pages 252–266.
30. S. Qadeer and J. Rehof. Context-Bounded Model Checking of Concurrent Software. In *TACAS'05*, pages 242–254.
31. V. Rehak. On Extensions of Process Rewrite Systems, PhD thesis, Masaryk University, 2007.
32. A. P. Sistla and E. M. Clarke. The Complexity of Propositional Linear Temporal Logics. *J. ACM* 32(3): 733-749 (1985)
33. W. Thomas. Constructing infinite graphs with a decidable MSO-theory. In *MFCS*, pages 113–124, 2003.
34. A. W. To and L. Libkin. Recurrent reachability analysis in regular model checking. In *LPAR'08*, pages 198–213.
35. M. Y. Vardi, P. Wolper. Automata-theoretic techniques for modal logics of programs. *JCSS* 32(2): 183–221 (1986).
36. K. N. Verma, H. Seidl, T. Schwentick. On the complexity of equational Horn clauses. In *CADE'05*, pages 337–352.